\documentclass[11pt]{article}
\usepackage{paspconf,graphics,amssymb,natbib,journals}

\newdimen\digitwidth
\setbox0=\hbox{\rm0}
\digitwidth=\wd0
\catcode`#=\active
\def#{\kern\digitwidth}
\begin{document}

\vglue -2cm
\noindent {\sl Contribution to IAU Colloq. 174 on Small Galaxy Groups, held in
Turku, FINLAND, June 13--18, 1999, ed. M. Valtonen \& C. Flynn, ASP series}

\vspace{1cm}

\title{Understanding low and high velocity dispersion compact groups}

\author{G. A. Mamon}
\affil{IAP, F-75014 Paris, FRANCE}





\begin{abstract}
A galaxy system must have a minimum velocity dispersion for its mass to be
greater than the sum of the masses of its galaxies.
Nearly half of the nearby Hickson compact groups (HCGs)
have too low a velocity dispersion in comparison with the
rotational velocities of their spiral galaxies and internal
velocity dispersions of their early types.

A detailed study of the
low velocity dispersion group, HCG~16 --- the only known
group of late-type galaxies 
with diffuse intergalactic X-ray emitting hot gas ---
reveals that half of the diffuse X rays are 
associated with foreground/background
sources and the remaining gas is 
clumpy and mostly associated with the bright galaxies of the group.
The large-scale environment of the group suggests that HCG~16 lies where a
cosmological filament falls perpendicularly onto a large-scale sheet.

The observed frequency of compact groups is lower than predicted from 
the extended Press-Schechter formalism, which also predicts that most $10^{13}
\,M_\odot$ objects in the Universe must be fairly old and hence have already
coalesced into single objects, reminiscent of elliptical galaxies
over-luminous in X-rays that are now being discovered.

Thus, the low survival time of dense groups against the merging instability
is no longer a worry for compact groups, as they form in large enough
numbers.  I show why other arguments against the reality of HCGs no longer
hold, partly because of the biases of Hickson's sample.

\end{abstract}




\section{Introduction}

Compact groups (hereafter CGs) have been puzzling astronomers
for a number of 
years.
How can a few bright galaxies coexist within 
less than 100 kpc?
CGs may have formed early, and have managed to survive the merging
instability (\citealp{GBC91}; Athanassoula, in these proceedings) 
or else formed just recently \citep{H82}.
Alternatively, CGs may be not be truly dense in 3D, but caused
instead by chance alignments of galaxies along the
line of sight within larger loose groups \citep{M86}, clusters \citep{WM89}
and cosmological filaments (\citealp{HKW95}, hereafter HKW).

In this contribution, we provide new light on this debate by studying the
group velocity dispersions, X-ray, optical and continuum radio emission, and
by predicting the frequency of dense groups as compact as
\citeauthor{H82}'s (\citeyear{H82}, hereafter HCGs) appear to be, using the
\citep{PS74} 
cosmological formalism. We conclude on the nature of HCGs.

\section{Low velocity dispersion compact groups}

\label{lowsigv}

For near spherical virialized systems, mass increases with some power (near
3, as is 
easily shown by combining the virial theorem with a critical mean density for
virialization) of the velocity dispersion of virialized systems.  One
therefore expects that \emph{there must be a minimum velocity dispersion for a
virialized galaxy system to be more massive than the sum of the masses of its
member galaxies}.
Systems near full collapse should have even larger velocity dispersions.

For a tighter constraint,
within a given radius $R$, the sum of the masses of the galaxies \emph{before
they they got close to one another} must be smaller or equal to the mass
within the same radius that
the group would have \emph{once it virializes}: $M(R) \geq \sum_j m_j(R)$.
Assuming \cite*{NFW95} profiles for groups and 
halos of
galaxies, one can show \citep{M99_lowsigv} that
the velocity dispersion $\sigma_v$ of a given dense group must
satisfy:
\begin{equation}
\sigma_v^2 \geq 0.16 \sum_{\rm spirals} 
v_{\rm rot}^2
+ 1.0 \sum_{\rm ellipticials} \sigma_{v,E}^2 
 \ ,
\label{maxveldsi}
\end{equation}
where the sums are over the deprojected maximum rotation velocities of
spirals and the internal velocity dispersions of ellipticals.

\begin{table}[h]
\caption{Hickson compact group minimum velocity dispersions}
\begin{center}\scriptsize
\begin{tabular}{lccccccc}
Group & $N_S$ & $\sigma_v^S$ & $N_E$ & $\sigma_v^E$ & $\sigma_v^{\rm min}$ &
$N$ & $\sigma_v$ \\
\tableline
HCG 16 & 4 & 151/166 & 0 & --- & 151/166 & 4 & #99 \\
HCG 23 & 2 & 123 & 0 & --- & 123 & 4 & 180 \\
HCG 33 & 1 & #93 & 0 & --- & #93 & 4 & 172 \\
HCG 34 & 1 & #83 & 0 & --- & #83 & 4 & 365 \\
HCG 37 & 1 & #93 & 2 & 266 & 282 & 5 & 445 \\
HCG 40 & 3 & 124 & 1 & 199 & 234 & 5 & 160 \\
HCG 44 & 2 & #98 & 1 & 158 & 186 & 4 & 145 \\
HCG 57 & 3 & 166 & 2 & 195 & 256 & 7 & 275 \\
HCG 79 & 1 & #42 & 0 & --- & #42 & 4 & 150 \\
HCG 88 & 2 & 118 & 0 & --- & 118 & 4 & #24 \\
HCG 89 & 2 & #79 & 0 & --- & #79 & 4 & #35 \\
HCG 90 & 3 & 106 & 0 & --- & 106 & 4 & 108 \\
\end{tabular}
\end{center}

\noindent Columns (2) and (4): number of spirals and ellipticals used;
columns (3) and (5): minimum group velocity dispersion (in $\, \rm km \,
s^{-1}$) from spirals and from
ellipticals; column (6): global minimum group velocity dispersion
($\rm km \, s^{-1}$), {\it
i.e.\/,} 
$\left (\sigma_v^{\rm min}\right)^2 = 
\left (\sigma_v^{\rm S}\right)^2 +
\left (\sigma_v^{\rm E}\right)^2$; column (7)
number of accordant redshift galaxies in group; column (8) measured group
velocity dispersion ($\rm km \, s^{-1}$).
\end{table}

Although there is still little data on the internal kinematics of HCG
galaxies, Table 1 above shows that 5 HCGs out of 12 (HCGs 16, 40, 44, 88 and
89) have abnormally low
velocity dispersions, and two others (HCGs 57 and 90)
have just marginal velocity dispersions.
We expect to measure by chance low velocity
dispersions in roughly 20\% of dense groups (those with chance tangential
velocity vectors). Still,
most of the 5 HCGs mentioned above have too low velocity dispersions to be
dense systems near virialization. The simplest alternative is that they 
are chance alignments within loose groups
near turnaround, since at turnaround the velocity dispersion of galaxy systems
is expected to be small.
Another possibility is that \emph{tidal friction} has been effective in
slowing down 
the galaxies, although this is only expected in much more compact
groups near full coalescence
(\citealp{WTK99}; Temporin, in these proceedings).

The two groups with $\sigma_v \simeq \sigma_v^{\rm min}$
may simply have little intergalactic matter.
Given that the internal kinematics data on HCG galaxies is very incomplete
(only one group has data for all its members, {\it i.e.\/,} $N = N_S+N_E$),
some of the groups with high velocity dispersions may turn out to have only
marginal velocity dispersions and thus possess little intergalactic matter,
while some of the marginal ones may in fact be non-real.
In any event, \emph{There seems to be 3 classes of compact groups, following
decreasing velocity dispersion: groups with substantial intergalactic matter,
groups will little intergalactic matter and chance alignments within loose
groups (or clusters or cosmological filaments)}.

Finally one may be tempted to secure more precise velocity dispersions by
including the galaxies from the environment of HCGs with the data of
\cite{dCRZ94} and \cite{ZM98art}, but these spectroscopic surveys show that
the velocity dispersion usually
increases with inclusion of the environment galaxies, and it is not clear
that one is not increasingly affected by interlopers.

\section{HCG 16}

We now focus on HCG~16, the prime, example of a low velocity dispersion,
spiral-rich compact group.

\subsection{X-ray emission}

The X-ray properties of HCG~16 are controversial and possibly
extreme.
In their ROSAT/PSPC X-ray survey of HCGs, 
\cite{PBEB96} (hereafter PBEB), HCG~16 was  
the coldest detected group ($T = 0.30\pm 0.05\,\rm
keV$), and there are no other spiral-only compact groups with diffuse X-ray
emission 
(Mulchaey in these proceedings; see also PBEB).
Moreover, whereas diffuse X-rays were clearly
detected by PBEB, \cite{SC95} failed
to detect such diffuse emission at an upper 
limit 16 times 
lower,\footnote{Given the fluxes measured by \cite{SC95}
for HCG~16 and their adopted value for $H_0$, their quoted upper 
limit for their
luminosities were underestimated by a factor 2 for all undetected
groups in their Table~4 except HCG 3.}
whereas only a factor 2.3
is attributable to the wider (``bolometric'') energy range
in which
PBEB compute their luminosities.
Given the low temperature that PBEB derive for HCG~16,
their derived X-ray luminosity places it two orders of
magnitude above their compact group
luminosity-temperature relation and roughly a factor of two above  the
extrapolation of the cluster trend.
It thus seems difficult to reconcile HCG~16 with a low temperature
extrapolation of regular X-ray emitting compact groups.

\begin{figure}[ht]
\begin{center}
\rotatebox{-90}{\resizebox{!}{0.49\hsize}{\includegraphics{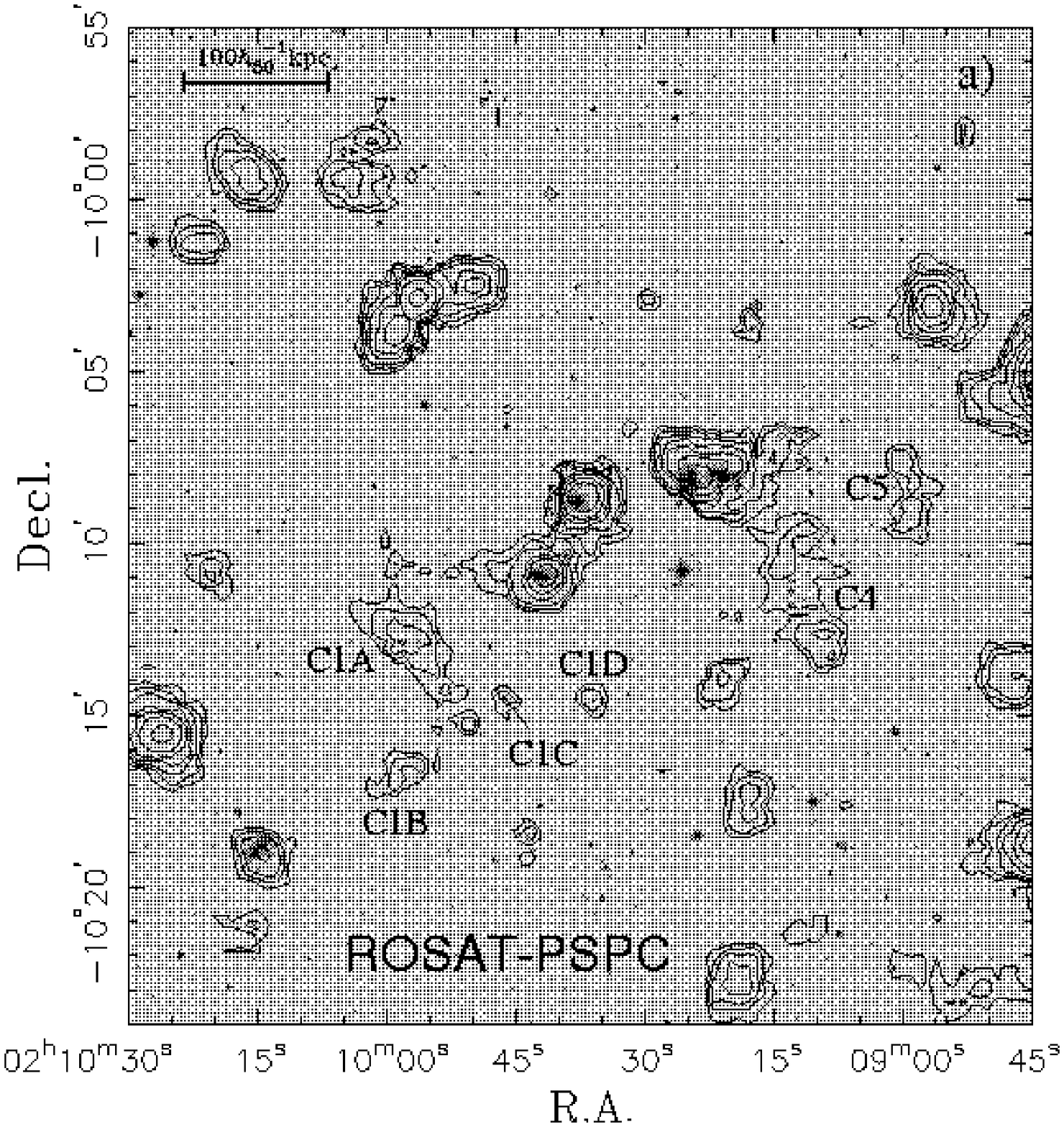}}}
\hfill\rotatebox{-90}{\resizebox{!}{0.49\hsize}{\includegraphics{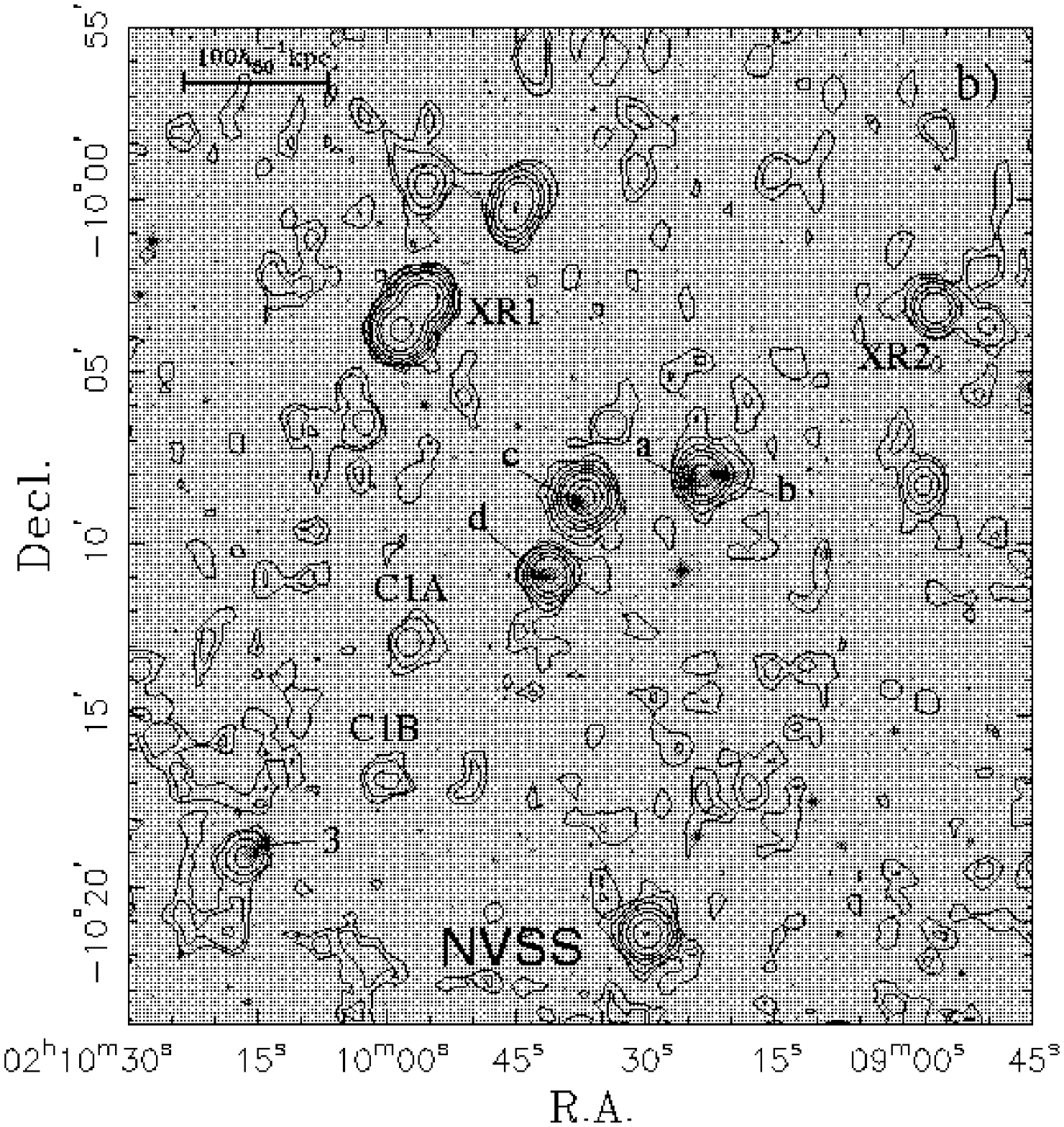}}}
\end{center}
\caption{Contour maps of (a): the adaptively smoothed (50 counts per
smoothing circle) {\sf ROSAT/PSPC} X-ray emission, b): the {\sf NVSS} 
20 cm radio emission of HCG~16,
both
superimposed on an optical {\sf DSS} image. 
The five polygonal regions (C1--C5)
dividing 
the emission region in HCG~16 are also shown, as well as the different
components of region C1 (C1A, C1B, C1C and C1D, see text).
}
\label{H16}

\end{figure}

\cite{DSM99} have re-analyzed the ROSAT/PSPC observations of HCG~16
with the hopes of resolving the discrepancy between \citeauthor{SC95} and
PBEB and establishing if an irregular morphology is
caused by clumpiness or fluctuations in signal-to-noise ratios.
Figure~\ref{H16}a below shows the X-ray emission of HCG~16 as contours
overlayed on a greyscale map of the group.
The emission beyond the galaxies is indeed significant in a few compact
regions: around the galaxies a\&b, c and d, as well as a few clumpy regions
outside the galaxies denoted C1 (comprised of 4 sub-clumps), C4 and C5.

Figure~\ref{H16}b shows the 20 cm continuum 
radio contours, measured with the {\sf NVSS}
survey, and illustrates \emph{the similarity between the X-ray and 20~cm 
continuum radio morphologies of HCG~16}.
A closer look \citep{DSM99} reveals that
C1A is connected with a radio-galaxy, which turns out to have a redshift
\citep{RdC3Z96} that
clearly places it in the background, C1B is connected with a radio-source,
C1C is related to a foreground star, and C5 is connected to a radio-source and
to a background group or cluster.
This reduces the X-ray luminosity of the diffuse hot gas connected with
HCG~16 by 50\% to $L_X^{\rm bol} = 2.3 \times
10^{41}\,h_{50}^{-2} \rm
erg\,s^{-1}$.
In the regions without significant X-ray
emission, the upper limits to the counts correspond to at most 1/4 the
the space density of hot gas in the detected regions.
Hence, \emph{the hot gas in HCG~16 is clumpy}.

\subsection{Large-scale environment}

If HCG~16 is part a cosmological filament viewed nearly end-on (HKW),
one should see this filament in its large-scale environment.
We have searched with {\sf NED} a roughly cubical region around HCG~16 of $40
\, h_{50}^{-1} \, \rm Mpc$ size.
HCG~16 is close enough ($cz = 3899 \, \rm km \, s^{-1}$)
that {\sf NED} should be fairly complete around it.

\begin{figure}[ht]
\begin{center}
\resizebox{0.49\hsize}{!}{\includegraphics{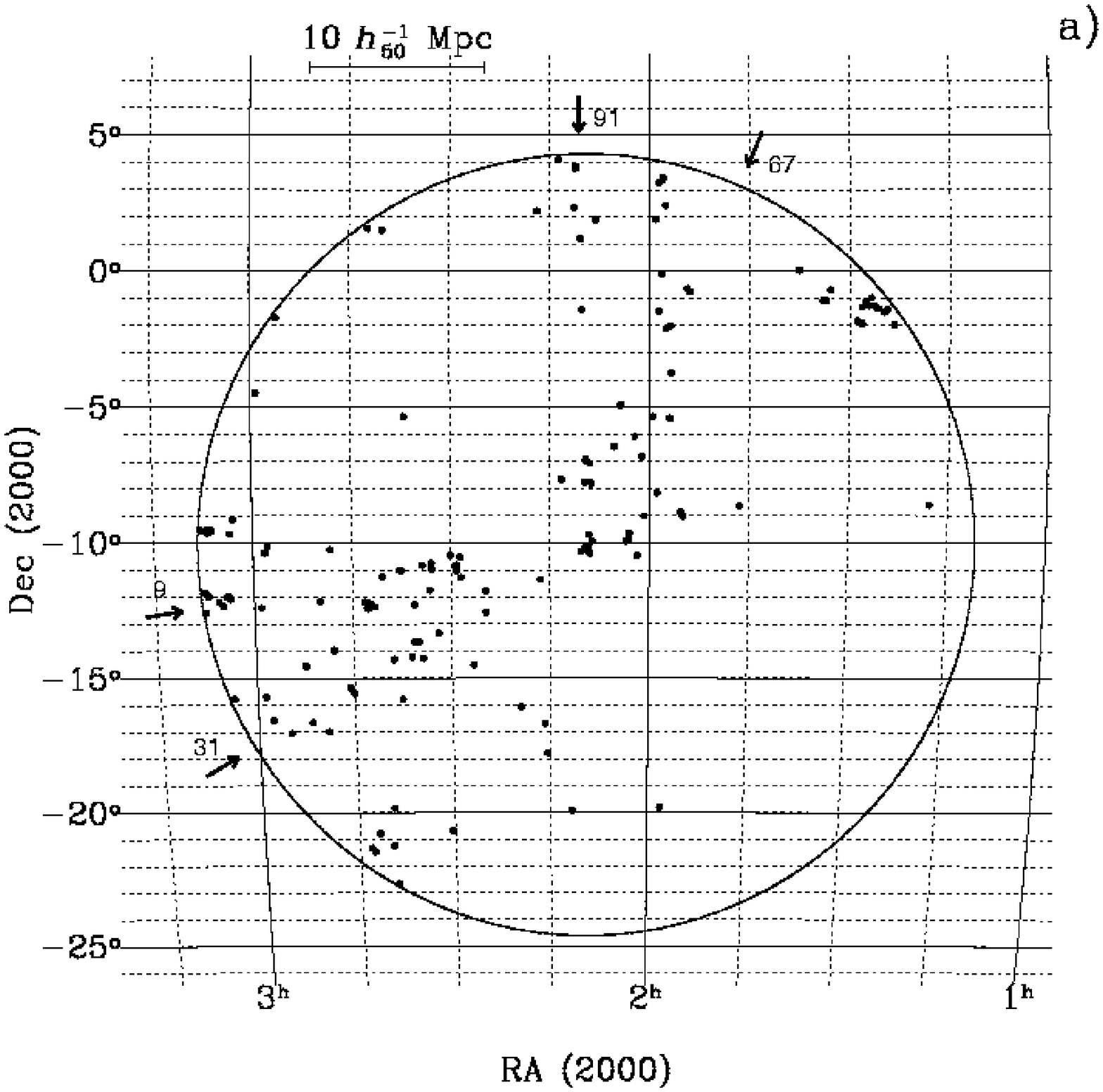}}
\resizebox{0.49\hsize}{!}{\includegraphics{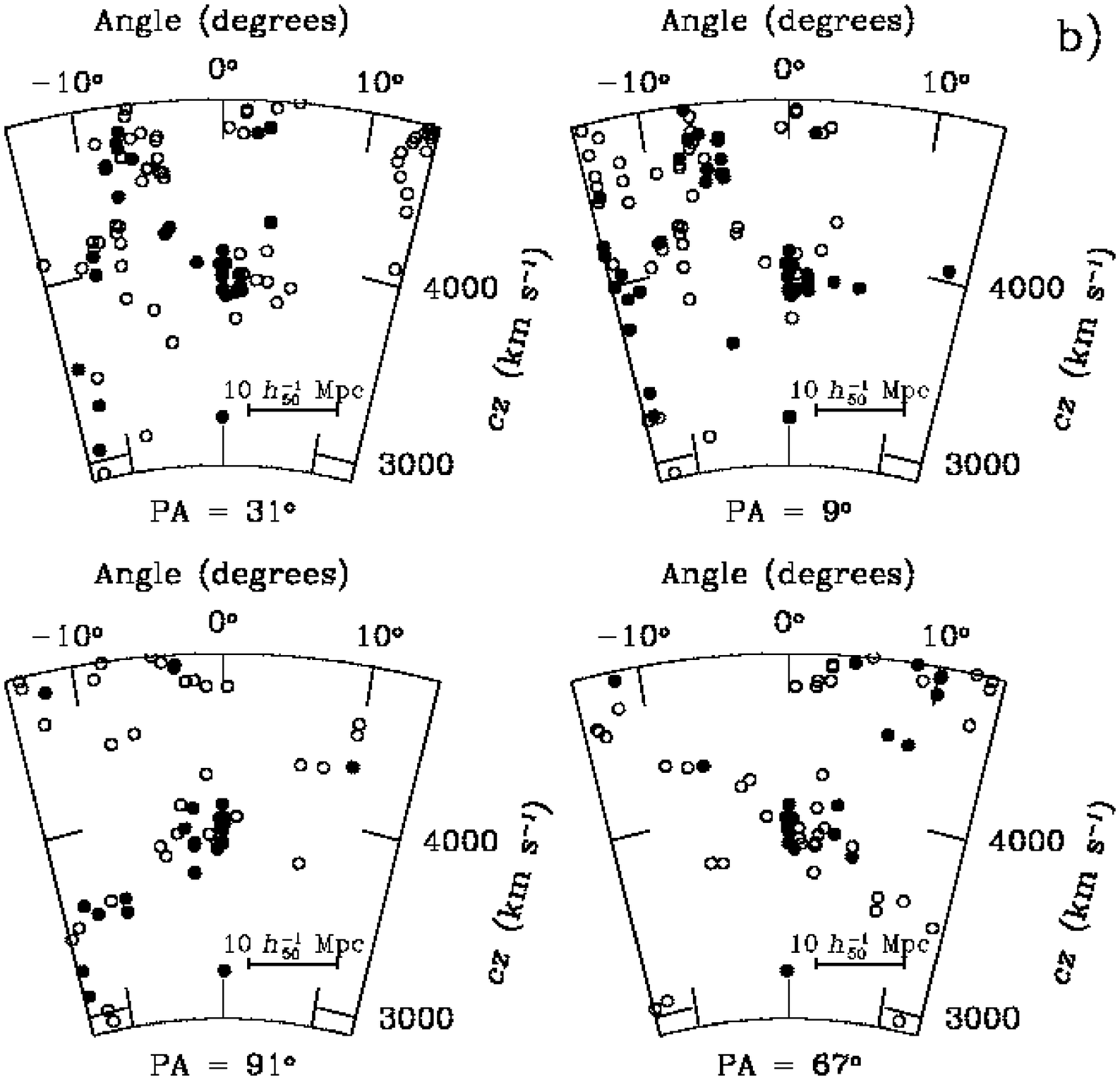}}
\end{center}
\caption{Environment of HCG~16, limited to $cz = 3899\pm1000 \, \rm km \,
s^{-1}$. a) Projected environment (HCG~16 in center).
b) Line-of-sight environments from wedge diagrams at 4
position angles. \emph{Filled} and \emph{open} symbols refer to galaxies
within projected 
distances of respectively 1 and $5 \,
h_{50}^{-1} \, \rm Mpc$ from the major axis of the projected wedge.
HCG~16 and its 3 neighboring galaxies appear as the \emph{finger of God}
of filled symbols at the center of each diagram.}
\label{envmt}
\end{figure}

The projected environment of HCG~16 within $\pm
1000 \, \rm km \, s^{-1}$ from the group distance (Fig.~\ref{envmt}a)
suggests concentration of
galaxies along 4 projection angles.
Figure \ref{envmt}b shows the wedge diagrams in each of these position
angles, and one can guess a filament at $\rm PA = 31^\circ$, stopping at
HCG~16 and a wide sheet
visible at $\rm PA = 91^\circ$ and $67^\circ$.
\emph{There are no filaments closely aligned to the line-of-sight} as would have
been favored by HKW.


\section{Cosmological predictions on the mass functions of groups}

\label{PS}

Although loose groups have generally not yet collapsed, one can apply the
\citeauthor{PS74} (\citeyear{PS74}, hereafter, PS) formalism to these
systems, assuming that when they will collapse, their mass is what we infer
today ($t=t_0$).
We then obtain
\begin{equation}
N_{\rm LG} (M,t_0) = \int_{t_0}^{2\,t_0} dt \,R_{\rm form} (M,t) \ ,
\end{equation}
where $R_{\rm form}$ is the rate of formation of structures derived by
\cite{KS96_PSdot} from the PS formalism.

$N$-body simulations suggest that compact groups cannot survive (with at
least 4 members) for over $\Delta t = 0.05-0.10\,t_0$ (\citealp{BCL93,AMB97}; 
Athanassoula in these proceedings),
the fraction increasing
with mass.
Assume therefore that compact groups must have undergone their cosmological
collapse within that time.
One then obtains
\begin{equation}
N_{\rm DG} (M,t_0) = \int_{t_0-\Delta t}^{t_0} dt \int_{M/2}^M 
dM' R_{\rm form} (M',t) P(M,t_0|M',t) \ ,
\end{equation}
where $P$ is the probability that a dense group of mass $M$ exists today 
given that it collapsed with a mass $M' < M$ at time $t < t_0$
(given by \citealp{LC93}).

\begin{figure}[ht]
\begin{center}
\resizebox{0.72\hsize}{!}{\includegraphics{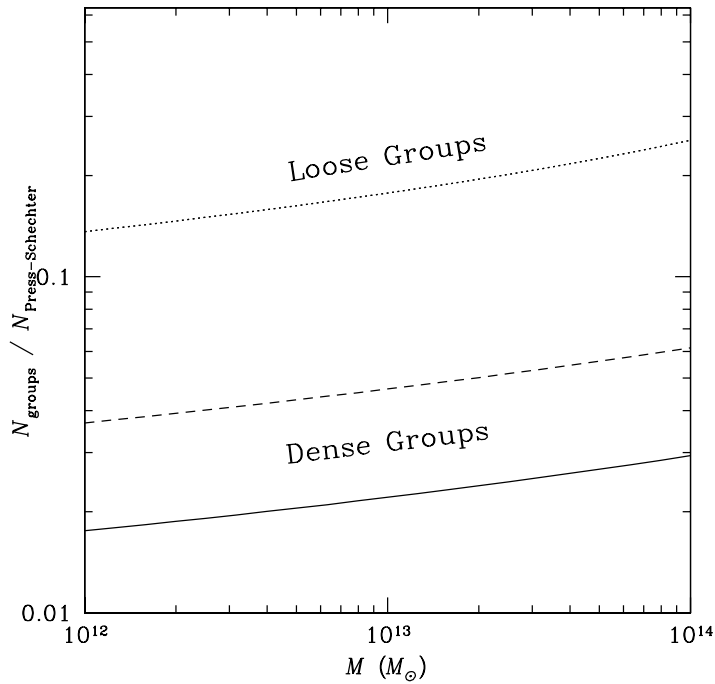}}
\end{center}
\caption{Loose ({\it thin dotted curve\/}) and dense ({\it thick 
curves\/}) group mass functions scaled to the Press-Schechter cosmic mass
function, derived from extended Press-Schechter theory (eqs.[2] and [3]),
assuming a $\Lambda$CDM cosmology with $\sigma_8 = 0.9$. {\it Dashed\/} and
{\it solid curves\/} refer to dense groups collapsing later than 0.9 and
$0.95\,t_0$, respectively.} 
\label{ndglgps}
\end{figure}

Figure~\ref{ndglgps} shows the resulting loose and dense group mass
functions, both normalized to the PS cosmic mass function, for a $\Lambda$CDM
cosmology (SCDM and OCDM with reasonable values for $\sigma_8$ yield roughly
similar curves).
Massive dense groups survive longer simply because they have more galaxies,
so 
the figure should be interpreted by adopting the recent collapse ({\it
solid curve\/}) for the lower mass end and the earlier collapse for the high
mass end.
The ratio of dense to loose groups then varies with increasing group
mass from 12 to 20\%. 
If compact groups are replenished by infalling galaxies \citep{GTC96}, they
collapse even earlier and we would predict even more compact groups
today.

In
contrast, the estimates of the ratio of HCGs to loose groups range from
0.5--8\% (\citealp{WM89}; see also \citealp{M86, M92_DAEC}).
Comparison of the HCG sample with a
similar compact group sample automatically selected from 
COSMOS/EDSGC galaxies (\citealp{PIM94}; Iovino in these proceedings) shows that
the HCG sample is severely incomplete: inspection of \citeauthor{PIM94}'s
Fig.~7 suggests that the incompleteness of the
HCG sample is a factor of 3
at the bright-end and increasingly worse at fainter magnitudes.

Now if dense groups represent only a few percent of the cosmic mass function,
where are the remaining virialized cosmic structures with masses near
$10^{13}\,M_\odot$?
The answer is simple: \emph{the remaining structures collapsed too early to be
visible as compact groups today, and must then be in different stages of the
final coalescence of groups}.
This includes quartets with massive dominant members (thus failing
\citeauthor{H82}'s \citeyear{H82} magnitude concordance criterion: $m-m_1 <
3$), as 
well as 
triplets, and binaries.
But since the cosmic multiplicity function decreases with increasing
galaxy number, one
is led to conclude that most of the missing structures are fully coalesced,
single galaxies, which harbor large X-ray halos, as discovered by
\citeauthor{MZ99} (\citeyear{MZ99}, see Mulchaey, in these proceedings) and
\cite{Vikhlinin+99}.
Therefore, we predict that \emph{X-ray over-luminous ellipticals should be
more common than compact groups}.

\section{Why Hickson's compact groups are dense in 3D after all}

It is very difficult to decide whether the numerous signatures of
galaxy interactions in CGs are proofs of their reality or alternatively 
simply binary interactions within binary-rich chance alignments of galaxies
within the line of sight \citep{M92_DAEC,HKW95}.
The high fraction of HCGs with diffuse X-ray emission (PBEB) should be
reduced when only considers groups with \emph{regular} 
diffuse X-rays (\S\ 3.1).
However, the combination of HI (Verdes-Montenegro, in these proceedings) and
optical 
(Mendes de Oliveira, in these proceedings) data on HCGs suggests that most of
these are real interacting systems. 
In view of this beautiful data, the arguments  spelled out by \cite{M86}
against the reality of HCGs 
must be reappraised.
 
The large predicted cosmic rate of production of dense groups (\S\,4
above) implies that the low survival time of dense groups is no longer a good
argument against the reality of most compact groups.
Moreover, the end products of dense groups --- isolated bright elliptical
galaxies with huge X-ray halos --- are now being seen
\citep{MZ99,Vikhlinin+99}.

Whereas mergers tend to increase rapidly the difference in first and second
ranked magnitudes, 
the low $\langle m_2-m_1\rangle$ of HCGs \citep{M86,M87} is simply caused by
\citeauthor{H82}'s strong
bias \citep{PIM94}
against groups with dominant members still satisfying
the magnitude concordance criterion.
This bias probably also causes the absence of significant luminosity
segregation in comparison with simulated groups \citep{M86,M87}.

The morphology-density relation of HCGs is
offset \citep{M86} relative to the cluster / loose
group / field morphology-density relation (measured by \citealp{PG84}):
at a given galaxy number density, HCGs have too many spirals.
However, HCGs have too \emph{few} spirals given their velocity dispersions
in comparison with the cluster morphology-velocity dispersion trend (PBEB).

There happens to be virtually no low velocity dispersion HCGs beyond 10,$000
\, \rm km \, s^{-1}$, therefore the number of HCGs whose low velocity
dispersion (\S\,2)
suggests they are non-real is limited to roughly 10 out of 69
accordant redshift HCGs with at least 4 members.

Therefore, \emph{there are few arguments left against the reality of HCGs}.
However, the samples of automatically defined compact groups (Iovino,
in these proceedings) will be much looser on average (given the same
compactness criterion as originally used by \citeauthor{H82}), since the HCG
sample is severely incomplete for 
marginally compact groups \citep{WM89,PIM94}, hence one would expect that
\emph{automatically selected compact groups will be more
prone to chance alignments} (see \citealp{WM89}).

\medskip

\acknowledgments

The work on HCG~16  was performed in collaboration with Sergio Dos
Santos. I thank Trevor Ponman for useful discussions.


\begin{thebibliography}{25}
\expandafter\ifx\csname natexlab\endcsname\relax\def\natexlab#1{#1}\fi
\itemsep 3pt
\bibitem[{Athanassoula} et~al.(1997){Athanassoula}, {Makino}, \&
  {Bosma}]{AMB97}
{Athanassoula}, E., {Makino}, J., {Bosma}, A., 1997, \mnras 286, 825

\bibitem[{Bode} et~al.(1993){Bode}, {Cohn}, \& {Lugger}]{BCL93}
{Bode}, P.W., {Cohn}, H.N., {Lugger}, P.M., 1993, \apj 416, 17

\bibitem[{de Carvalho} et~al.(1994){de Carvalho}, {Ribeiro}, \& {Zepf}]{dCRZ94}
{de Carvalho}, R.R., {Ribeiro}, A.L.B., {Zepf}, S.E., 1994, \apjs 93, 47

\bibitem[{Dos Santos} \& {Mamon}(1999)]{DSM99}
{Dos Santos}, S., {Mamon}, G.A., 1999, \aap in press, astro-ph/9811271

\bibitem[{Governato} et~al.(1991){Governato}, {Bhatia}, \& {Chincarini}]{GBC91}
{Governato}, F., {Bhatia}, R., {Chincarini}, G., 1991, \apjl 371, L15

\bibitem[{Governato} et~al.(1996){Governato}, {Tozzi}, \& {Cavaliere}]{GTC96}
{Governato}, F., {Tozzi}, P., {Cavaliere}, A., 1996, \apj 458, 18

\bibitem[{Hernquist} et~al.(1995){Hernquist}, {Katz}, \& {Weinberg}]{HKW95}
{Hernquist}, L., {Katz}, N., {Weinberg}, D.H., 1995, \apj 442, 57 (HKW)

\bibitem[{Hickson}(1982)]{H82}
{Hickson}, P., 1982, \apj 255, 382

\bibitem[{Kitayama} \& {Suto}(1996)]{KS96_PSdot}
{Kitayama}, T., {Suto}, Y., 1996, \mnras 280, 638

\bibitem[{Lacey} \& {Cole}(1993)]{LC93}
{Lacey}, C., {Cole}, S., 1993, \mnras 262, 627

\bibitem[{Mamon}(1986)]{M86}
{Mamon}, G.A., 1986, \apj 307, 426

\bibitem[{Mamon}(1987)]{M87}
{Mamon}, G.A., 1987, \apj 321, 622

\bibitem[{Mamon}(1992)]{M92_DAEC}
{Mamon}, G.A., 1992, in 2nd DAEC mtg.,
  Distribution of Matter in the Universe, G.A. {Mamon}, D. {Gerbal}, Paris: Obs. de Paris,  p. 51, ftp://ftp.iap.fr/pub/from\_users/gam/PAPERS/daec92-cg.dvi.Z

\bibitem[{Mamon}(1999)]{M99_lowsigv}
{Mamon}, G.A., 1999, to be submitted to \aap

\bibitem[{Mulchaey} \& {Zabludoff}(1999)]{MZ99}
{Mulchaey}, J.S., {Zabludoff}, A.I., 1999, \apj 514, 133

\bibitem[{Navarro} et~al.(1995){Navarro}, {Frenk}, \& {White}]{NFW95}
{Navarro}, J.F., {Frenk}, C.S., {White}, S.D.M., 1995, \mnras 275, 720

\bibitem[{Ponman} et~al.(1996){Ponman}, {Bourner}, {Ebeling}, \&
  {B\"ohringer}]{PBEB96}
{Ponman}, T.J., {Bourner}, P.D.J., {Ebeling}, H., {B\"ohringer}, H., 1996,
  \mnras 283, 690 (PBEB)

\bibitem[{Postman} \& {Geller}(1984)]{PG84}
{Postman}, M., {Geller}, M.J., 1984, \apj 281, 95

\bibitem[{Prandoni} et~al.(1994){Prandoni}, {Iovino}, \& {MacGillivray}]{PIM94}
{Prandoni}, I., {Iovino}, A., {MacGillivray}, H.T., 1994, \aj 107, 1235

\bibitem[{Press} \& {Schechter}(1974)]{PS74}
{Press}, W.H., {Schechter}, P., 1974, \apj 187, 425

\bibitem[{Ribeiro} et~al.(1996){Ribeiro}, {de Carvalho}, {Coziol}, {Capelato},
  \& {Zepf}]{RdC3Z96}
{Ribeiro}, A.L.B., {de Carvalho}, R.R., {Coziol}, R., {Capelato}, H.V., {Zepf},
  S.E., 1996, \apjl 463, L5

\bibitem[{Saracco} \& {Ciliegi}(1995)]{SC95}
{Saracco}, P., {Ciliegi}, P., 1995, \aap 301, 348

\bibitem[{Vikhlinin} et~al.(1999){Vikhlinin}, {McNamara}, {Hornstrup},
  {Quintana}, {Forman}, {Jones}, \& {Way}]{Vikhlinin+99}
{Vikhlinin}, A., {McNamara}, B.R., {Hornstrup}, A., {Quintana}, H., {Forman},
  W., {Jones}, C., {Way}, M., 1999, \apjl 520, L1

\bibitem[{Walke} \& {Mamon}(1989)]{WM89}
{Walke}, D.G., {Mamon}, G.A., 1989, \aap 225, 291

\bibitem[{Weinberger} et~al.(1999){Weinberger}, {Temporin}, \& {Kerber}]{WTK99}
{Weinberger}, R., {Temporin}, S., {Kerber}, F., 1999, \apj in press,
  astro-ph/9907304

\bibitem[{Zabludoff} \& {Mulchaey}(1998)]{ZM98art}
{Zabludoff}, A.I., {Mulchaey}, J.S., 1998, \apj 496, 39

\end{thebibliography}
\end{document}